# Constrained Forms of the Tsallis Entropy Function and Local Equilibria

Robert K. Niven[1]

[1]School of Aerospace, Civil and Mechanical Engineering, The University of New South Wales at ADFA, Northcott Drive, Canberra, ACT, 2600, Australia. Email: r.niven@adfa.edu.au

**Abstract**

The Lagrangian technique of Niven (2004, *Physica A*, 334(3-4): 444) is used to determine the *constrained* forms of the Tsallis entropy function - i.e. Lagrangian functions in which the probabilities of each state are independent - for each constraint type reported in the literature (here termed the Mark I, II and III forms). In each case, a constrained form of the Tsallis entropy function exists, which at $q=1$ reduces to its Shannon equivalent. Since they are fully constrained, each constrained Tsallis function can be "dismembered" to give its *partial* or *local* form, providing the means to independently examine each state $i$ relative to its local stationary (maximum entropy) position. The Mark II and III functions depend on $q$, the probability, the stationary probability, and the respective $q$-partition function; in contrast the Mark I form depends only on the first three parameters. The Mark II and III forms therefore depend on the structure of the system. The utility of the dismemberment method is illustrated for a system with equispaced energy levels.

*Keywords:* information theory; Tsallis entropy; constrained; statistical mechanics; non-extensive.

1. **Introduction**

Over the past 15 years, a major advance has taken place in theoretical physics, with the development of "non-extensive" (more accurately, long-range correlated) statistical mechanics based on the Tsallis entropy function [1]. A superset of "extensive" or Boltzmann-Gibbs statistical mechanics, it provides a method to determine the "most probable" or maximum entropy position of systems in which the probabilities (and entropies) associated with systems A and B become correlated when combined to form system AB [1-4]. Although still in its infancy, Tsallis' statistical mechanics has been invoked in the analysis of a diverse range of complicated systems, including those with long-range interactions, long-range memory effects and multifractal (power-law) space-time structures [3-4]. Examples of particular interest include Levy-type and correlated-type anomalous diffusion [5-6], fluid and financial turbulence [7] and fluidized granular systems [8].

Recently, Niven [9] introduced the *constrained* or Lagrangian forms of the Shannon entropy and Kullback-Leibler cross-entropy functions, which incorporate the constraints on the system, irrespective of the nature (or even the number) of the constraints. Since they are fully constrained, their summations can be "dismembered" to give their *partial* or *local* form, providing the means to independently examine each outcome or state $i$ relative to its local maximum entropy (or minimum cross-entropy) position [9]. Such functions are suitable for the examination of systems within the framework of Boltzmann-Gibbs statistical mechanics.

The aim of the present study is to apply the constraining method to the Tsallis entropy function, to determine its partial or local form under various conditions [3-4]. In Part 2 a brief account of background theory is provided, leading in Part 3 to the derivation of the constrained forms of the Tsallis entropy function. The utility of the method is illustrated in Part 4, for an example system with equispaced energy levels.

## 2. Background

### 2.1 *The Shannon Entropy and Constraining Method*

The discrete *information entropy function* of Shannon [10] is, in dimensionless form:

$$H = \frac{S}{k} = -\sum_{i=1}^{s} p_i \ln p_i, \tag{1}$$

where $p_i$ is the probability of occurrence of the $i$th outcome or state, $s$ is the number of possible outcomes or states, $S$ is the dimensional Shannon entropy function and $k$ is the Boltzmann or other scaling constant. Usually eq. (1) is subject to the *natural constraint* and one or several *moment constraints*, respectively:

$$\sum_{i=1}^{s} p_i = 1, \tag{2}$$

$$\sum_{i=1}^{s} p_i F_{ri} = \langle F_r \rangle, \quad r = 1,...,R, \tag{3}$$

where $F_{ri}$ is the value of the function or observable $F_r$ in the $i$th state and $\langle F_r \rangle$ is the mathematical expectation of $F_r$. Applying the Lagrange method of undetermined multipliers to the Shannon entropy (Eq. (1)) subject to the constraints (Eqs. (2) and (3)) yields the Lagrangian:

$$\mathcal{L} = \sum_{i=1}^{s} \left\{ -p_i \ln p_i - (\lambda_0 - 1)p_i - \sum_{r=1}^{R} \lambda_r p_i F_{ri} \right\}, \qquad (4)$$

where $\lambda_0$, $\lambda_r$, $r=1,...,R$, are the Lagrangian multipliers, and $(\lambda_0 - 1)$ is chosen for convenience. Extremisation of Eq. (4) (by setting $d\mathcal{L}/dp_i = 0$, $\forall i$) gives the maximum entropy (equilibrium) distribution $p_i$* as the *generalised Maxwell-Boltzmann distribution* of statistical mechanics [11-12]:

$$p_i^* = \exp\left(-\lambda_0 - \sum_{r=1}^{R} \lambda_r F_{ri}\right) = \frac{1}{Z} \exp\left(-\sum_{r=1}^{R} \lambda_r F_{ri}\right), \quad i = 1,...,s, \qquad (5)$$

where $Z = \sum_{i=1}^{s} \exp\left(-\sum_{r=1}^{R} \lambda_r F_{ri}\right)$ is the generalised partition function.

In an interesting analysis by Niven [9], the constraints in the Lagrangian (Eq. (4)) may be replaced by the equilibrium probability $p_i$*, using the first form of Eq. (5), giving the *constrained* form of the Shannon entropy function:

$$\mathcal{L} = H^c = \sum_{i=1}^{s} -p_i \ln \frac{p_i}{p_i^*} + p_i . \qquad (6)$$

$H^c$ is a generic version of the Shannon entropy function, valid for a closed system irrespective of the form or even the number of the constraints.

Since $H^c$ is a Lagrangian, it is fully constrained; i.e. all $p_i$ are independent. The summation may therefore be "dismembered" to give the *partial constrained form* of the

Shannon entropy [9]:

$$H_i^c = -p_i \ln\left(\frac{p_i}{p_i{*}}\right) + p_i, \quad i=1,\ldots,s. \tag{7}$$

The significance of $H_i^c$ is that it enables each state $i$ to be examined relative to its *local* equilibrium position, independently of the other states in the system. The partial form of the Shannon entropy, $H_i = -p_i \ln p_i$, does not possess this functionality. Thus although $H^c$ appears to be a simple variant of the Kullback-Leibler cross-entropy, $D = \sum_{i=1}^{s} p_i \ln(p_i/\pi_i)$, where $\pi_i$ are the *a priori* probabilities, in which we have set $\pi_i = p_i{*}, \forall i$, we are here interested in the *local* optimum of each state, and not simply the global optimum. Accordingly, each $H_i^c$ must include the second positive $p_i$ term [9].

A three-dimensional plot of $H_i^c$ against $p_i$ and $p_i{*}$ is shown in Fig. 1 (after [9]). The properties of Eq. (7) and its continuous form are examined further elsewhere [9].

## 2.2 *The Tsallis Entropy*

The discrete *Tsallis entropy function* is [1-4]:

$$H_q = \frac{S_q}{k} = -\sum_{i=1}^{s} p_i^q \ln_q p_i = \frac{1}{q-1}\left(1 - \sum_{i=1}^{s} p_i^q\right), \tag{8}$$

where $S_q$ is the dimensional Tsallis entropy, $q \in \mathbb{R}$ is the Tsallis parameter, and $\ln_q f = (1-q)^{-1}(f^{1-q} - 1)$, $f>0$ is the $q$-logarithmic function. In the limit as $q \to 1$, $H_q \to H$. The Tsallis entropy is usually extremised subject to the normal constraint (Eq. (2)), and one or several moment constraints of three different types, referred to here as the Mark I, II and III forms [1-4]:

Mark I:
$$\sum_{i=1}^{s} p_i F_{ri} = \langle F_r \rangle, \quad r = 1,...,R, \quad (9)$$

Mark II:
$$\sum_{i=1}^{s} p_i^q F_{ri} = \langle F_r \rangle_q, \quad r = 1,...,R, \quad (10)$$

Mark III:
$$\frac{\sum_{i=1}^{s} p_i^q F_{ri}}{\sum_{i=1}^{s} p_i^q} = \langle\langle F_r \rangle\rangle_q, \quad r = 1,...,R, \quad (11)$$

where $\langle F_r \rangle_q$ and $\langle\langle F_r \rangle\rangle_q$ are, respectively, the unnormalised and normalised $q$-expectation of $F_r$. The Mark I form, identical to Eq. (3), was used in early literature, but suffers from mathematical difficulties such as undesirable diverging moments [3-4]. The Mark II form, proposed by Curado & Tsallis [2], has certain unexpected consequences such as non-additivity and non-translatability of energies [3-4]; also in general $\langle 1 \rangle_q \neq 1$. The Mark III form, using the *escort probabilities* $p_i^q / \sum_{i=1}^{s} p_i^q$ (whence $\langle\langle 1 \rangle\rangle_q = 1, \forall q$), is now preferred [3-4]. It should be noted that the Mark I, II or III (or some other) set of moment constraints give rise to different statistics. Apart from the above-mentioned mathematical niceties and certain experimental evidence for Mark III statistics, no *a priori* rationalisation has been given for any particular constraint set (nor whether there exist other constraint sets or mixed constraint problems).

The utility of the Tsallis entropy is that it forms the basis of a (fascinating) body of "non-extensive" statistical mechanics, applicable to systems in which there are significant long-range interactions between entities, in space and/or in time, with a diverse range of applications [3-4]. However, it would be mathematically useful to determine whether any constrained form(s) of the Tsallis entropy exist, which would enable each state $i$ to be examined independently of the other states. This study explores this issue further.

## 3. Forms of the Constrained Tsallis Entropy

### 3.1 Mark I Constraint Set

The Tsallis entropy (Eq. (8)) subject to the natural (Eq. (2)) and Mark I constraints (Eq. (9)) gives the Lagrangian (c.f. [1]):

$$\mathcal{L}^{(I)} = \frac{1}{q-1} + \sum_{i=1}^{s}\left\{-\frac{p_i^q}{q-1} - \kappa_0 p_i - \sum_{r=1}^{R}\kappa_r p_i F_{ri}\right\}, \qquad (12)$$

where $\kappa_0$, $\kappa_r$, $r=1,...,R$, are the Lagrangian multipliers, and superscript *(I)* denotes the Mark I form. Extremisation (of the summand) gives the stationary[1] or "most probable" distribution $p_i^{(I)}$:

$$p_i^{(I)} = \left[-\frac{q-1}{q}\left(\kappa_0 + \sum_{r=1}^{R}\kappa_r F_{ri}\right)\right]^{1/(q-1)}, \quad i=1,...,s. \qquad (13)$$

---

[1] In non-extensive statistical mechanics, the term "stationary" is preferred over "equilibrium" for the most probable (maximum entropy) distribution.

From the natural constraint (Eq. (2)), this reduces to:

$$p_i^{(I)} = \frac{1}{Z_q^{(I)}} \left(1 + \sum_{r=1}^{R} \kappa'_r F_{ri}\right)^{1/(q-1)}, \quad i = 1,...,s, \qquad (14)$$

where $Z_q^{(I)} = \sum_{i=1}^{s}(1 + \sum_{r=1}^{R}\kappa'_r F_{ri})^{1/(q-1)}$ is the $q$-partition function (Mark I), and $\kappa'_r = \kappa_r / \kappa_0$ is a modified Lagrangian multiplier. Eq. (14) with $\kappa'_r = -\beta(q-1)$ gives the form given by Tsallis ([1], Eq. (11)) when subject to a single (energy) moment constraint.

Applying the constraining method described, by substituting Eq. (13) into Eq. (12), yields the *constrained form (Mark I)* of the Tsallis entropy as:

$$\mathcal{L}^{(I)} = H_q^{c(I)} = \sum_{i=1}^{s} \frac{p_i \left[q \, (p_i^{(I)})^{q-1} - p_i^{q-1}\right]}{q-1}, \qquad (15)$$

from which the partial form is:

$$H_{q,i}^{c(I)} = \frac{p_i \left[q \, (p_i^{(I)})^{q-1} - p_i^{q-1}\right]}{q-1}, \quad i = 1,...,s. \qquad (16)$$

This is a function of only $p_i$, $p_i^{(I)}$ and $q$. Thus in the Mark I case, the constrained Tsallis entropy function exists, is independent of all other states, and apart from $p_i$ and $p_i^{(I)}$ is only dependent on $q$.

Interestingly, extremisation using an alternative form of the Tsallis entropy ($H_q = \sum_{i=1}^{s}(q-1)^{-1}(p_i - p_i^q)$, c.f. eq. (8)) yields a different form of $p_i^{(I)}$, but the same constrained form (Eq.(15)) is recovered.

Three-dimensional plots of $H_{q,i}^{c(I)}$ against $p_i$ and $p_i^{(I)}$ for various values of $q$ are illustrated in Figs. 2a-g. These can be compared to the $q=1$ case in Fig. 1. As evident, for $1<q<\infty$, $H_{q,i}^{c(I)}$ becomes increasingly saddle-like (convex with respect to $p_i^{(I)}$ and concave with respect to $p_i$) as $q\to\infty$. At the latter limit it collapses onto $H_{q,i}^{c(I)}=0$, except (peculiarly) at $p_i^{(I)}=1$, at which $H_{q,i}^{c(I)}=p_i$. In contrast, in the range $0<q\leq1$, $H_{q,i}^{c(I)}$ is fully concave, with enhanced concavity relative to $H_i^c$ as $q\to0^+$. At $q=0$, it collapses onto $H_{q,i}^{c(I)}=1$. For $q<0$, $H_{q,i}^{c(I)}$ is convex, becoming increasing so as $q\to-\infty$. In all cases $q\neq0$, the extremum occurs at $p_i = p_i^{(I)}$, at which $H_{q,i}^{c(I)} = (p_i^{(I)})^q$. The partial constrained Tsallis entropy (Mark I) therefore reflects the behaviour of each state $i$ relative to its local stationary position $p_i^{(I)}$. Note also that for $q>0$, $H_{q,i}^{c(I)}$ is negative over some of its range, a feature in common with $H_i^c$ [9].

### 3.2 *Mark II Constraint Set*

The Lagrangian of the Tsallis entropy (Eq. (8)) subject to the natural (Eq. (2)) and Mark II constraints (Eq. (10)) is (c.f. [2,3]):

$$\mathcal{L}^{(II)} = \frac{1}{q-1} + \sum_{i=1}^{s}\left\{-\frac{p_i^q}{q-1} - \mu_0 p_i - \sum_{r=1}^{R}\mu_r p_i^q F_{ri}\right\}, \quad (17)$$

where $\mu_0$, $\mu_r$, $r=1,...,R$, are the Lagrangian multipliers, and superscript *(II)* denotes the Mark II form. Extremisation gives the Mark II stationary distribution $p_i^{(II)}$ as:

$$p_i^{(II)} = \left[ \frac{\mu_0(1-q)}{q\left(1-(1-q)\sum_{r=1}^{R}\mu_r F_{ri}\right)} \right]^{1/(q-1)}, \quad i=1,...,s. \tag{18}$$

Again from Eq. (2):

$$p_i^{(II)} = \frac{1}{Z_q^{(II)}} \left(1-(1-q)\sum_{r=1}^{R}\mu_r F_{ri}\right)^{1/(1-q)} = \frac{1}{Z_q^{(II)}} \exp_q\left(-\sum_{r=1}^{R}\mu_r F_{ri}\right), \quad i=1,...,s, \tag{19}$$

where $Z_q^{(II)} = \sum_{i=1}^{s} \exp_q\left(-\sum_{r=1}^{R}\mu_r F_{ri}\right)$ is the $q$-partition function (Mark II), and $\exp_q f = [1+(1-q)f]^{1/(1-q)}$ is the $q$-exponential function. (Note other variants of Eq. (19) are possible, depending on the definition of $Z_q^{(II)}$.) Eq. (19) reduces to the form of Curado & Tsallis ([2], Eq. (10)) when subject to a single energy constraint.

Applying the constraining method to Eq. (18) gives the *constrained form (Mark II)* of the Tsallis entropy as:

$$\mathcal{L}^{(II)} = H_q^{c(II)} = \sum_{i=1}^{s} \frac{\mu_0 p_i \left[(p_i/p_i^{(II)})^{q-1} - q\right]}{q}. \tag{20}$$

From the foregoing extremisation, $\mu_0 = -q/[(q-1)(Z_q^{(II)})^{q-1}]$, whence:

$$\mathcal{L}^{(II)} = H_q^{c(II)} = \sum_{i=1}^{s} -\frac{p_i \left[(p_i/p_i^{(II)})^{q-1} - q\right]}{(q-1)(Z_q^{(II)})^{q-1}}. \tag{21}$$

The partial form $H_{q,i}^{c(II)}$ is thus:

$$H_{q,i}^{c(II)} = -\frac{p_i \left[(p_i/p_i^{(II)})^{q-1} - q\right]}{(q-1)(Z_q^{(II)})^{q-1}}. \tag{22}$$

As evident, the partial constrained entropy is not only a function of $p_i$, $p_i^{(II)}$ and $q$, but also of the $q$-partition function $Z_q^{(II)}$. Thus although the Mark II partial constrained entropy exists, it depends on the overall structure of the system - as expressed in $Z_q^{(II)}$ - and not just on the individual properties of state $i$. Such behaviour accords with what we might expect of non-extensive statistics, and stands in contrast to the Mark I form.

As $Z_q^{(II)}$ will not normally be known in advance, it is convenient to define a *scaled partial constrained form* (Mark II) of the Tsallis entropy as:

$$S_{q,i}^{c(II)} = H_{q,i}^{c(II)} (Z_q^{(II)})^{q-1} = -\frac{p_i \left[ (p_i / p_i^{(II)})^{q-1} - q \right]}{(q-1)}. \tag{23}$$

A (fascinating) feature of both $S_{q,i}^{c(II)}$ and $H_{q,i}^{c(II)}$ is that in the limit as $q \to 1$, both $S_{q,i}^{c(II)} \to H_i^c$ and $H_{q,i}^{c(II)} \to H_i^c$ (Eq. (6)), i.e. they both give the partial constrained form of the Shannon entropy [9]. Clearly, in Mark II non-extensive statistics ($q \neq 1$) it is necessary to take the $q$-partition function into account; however, in the limit as $q \to 1$, this is no longer necessary.

Plots of $S_{q,i}^{c(II)}$ against $p_i$ and $p_i^{(II)}$ for various values of $q$ are illustrated in Figs. 3a-g. Again the $q=1$ case is illustrated in Fig. 1. As expected, the extremum with respect to $p_i$, $\forall$ $q \neq 0$, occurs at $p_i = p_i^{(II)}$, at which $S_{q,i}^{c(II)} = p_i^{(II)}$ (n.b. at $q=0$, $S_{q,i}^{c(II)} = p_i^{(II)}$, $\forall p_i$, and there is no extremum). The Mark II partial constrained entropy may therefore be used to examine local stationary behaviour. As evident from the plots, for $q>0$ ($q<0$), $S_{q,i}^{c(II)}$ is concave (convex), becoming increasingly "folded" over the $p_i = p_i^{(II)}$ line as $q \to \infty$ ($q \to -\infty$). At $q=0$, as noted, it collapses onto the $S_{q,i}^{c(II)} = p_i^{(II)}$ plane. $S_{q,i}^{c(II)}$ has two zeros, at $p_i = 0$ and $p_i = q^{1/(q-1)} p_i^{(II)}$, the latter applying only for $q>0$. In consequence, for $q>0$, $S_{q,i}^{c(II)}$ can be negative over part of its range.

### 3.3 *Mark III Constraint Set*

For the Mark III constraint set (Eq. (11)), the original derivation of Tsallis [3] causes certain difficulties, which are overcome by expressing the constraints in the form given by Martínez *et al.* [13]:

$$\sum_{i=1}^{s} p_i^q \left( F_{ri} - \langle\langle F_r \rangle\rangle_q \right) = 0, \quad r = 1,...,R. \tag{24}$$

The Lagrangian of the Tsallis entropy (Eq. (8)) subject to Eqs. (2) and (24) is then:

$$\mathfrak{L}^{(III)} = \frac{1}{q-1} + \sum_{i=1}^{s} \left\{ -\frac{p_i^q}{q-1} - \nu_0 p_i - \sum_{r=1}^{R} \nu_r p_i^q (F_{ri} - \langle\langle F_r \rangle\rangle_q) \right\}, \tag{25}$$

where $\nu_0$, $\nu_r$, $r=1,...,R$, are the Lagrangian multipliers of Martínez *et al.* [13], and superscript (III) denotes Mark III statistics. Extremisation gives the stationary probability $p_i^{(III)}$ as:

$$p_i^{(III)} = \left[ \frac{\nu_0(1-q)}{q\left(1-(1-q)\sum_{r=1}^{R}\nu_r(F_{ri}-\langle\langle F_r \rangle\rangle_q)\right)} \right]^{1/(q-1)}, \quad i=1,...,s. \tag{26}$$

Applying the natural constraint (Eq. (2)) gives [13]:

$$p_i^{(III)} = \frac{1}{Z_q^{(III)}} \left( 1 - (1-q) \sum_{r=1}^{R} \nu_r (F_{ri} - \langle\langle F_r \rangle\rangle_q) \right)^{1/(1-q)}$$
$$= \frac{1}{Z_q^{(III)}} \exp_q \left( -\sum_{r=1}^{R} \nu_r (F_{ri} - \langle\langle F_r \rangle\rangle_q) \right), \quad i = 1, ..., s \tag{27}$$

where $Z_q^{(III)} = \sum_{i=1}^{s} \exp_q \left( -\sum_{r=1}^{R} \nu_r (F_{ri} - \langle\langle F_r \rangle\rangle_q) \right)$ is the Mark III $q$-partition function.

Applying the constraining method to Eq. (26) gives the *constrained form (Mark III)* of the Tsallis entropy as:

$$\mathcal{L}^{(III)} = H_q^{c(III)} = \sum_{i=1}^{s} \frac{\nu_0 p_i \left[ (p_i / p_i^{(III)})^{q-1} - q \right]}{q}. \tag{28}$$

Again from $\nu_0 = -q / [(q-1)(Z_q^{(III)})^{q-1}]$, we obtain:

$$\mathcal{L}^{(III)} = H_q^{c(III)} = \sum_{i=1}^{s} -\frac{p_i \left[ (p_i / p_i^{(III)})^{q-1} - q \right]}{(q-1)(Z_q^{(III)})^{q-1}}. \tag{29}$$

The partial form $H_{q,i}^{c(III)}$ is given by the summand of Eq. (29). Note Eqs. (28) and (29) are of the same form as the Mark II case (Eqs. (20) and (21)). As in that case, $H_{q,i}^{c(III)}$ is a function of $p_i$, $p_i^{(II)}$, $q$ and $Z_q^{(III)}$, and depends on the overall structure of the system as well as the properties of the local state.

As with the other Mark forms, extremisation using alternative formulations of the Lagrangian (eq. (25)) can give different forms of $p_i^{(III)}$, but the same constrained form (Eq. (28)) is recovered.

The scaled partial form of Eq. (29) is thus:

$$S_{q,i}^{c(III)} = H_{q,i}^{c(III)}(Z_q^{(III)})^{q-1} = -\frac{p_i\left[(p_i/p_i^{(III)})^{q-1} - q\right]}{(q-1)}. \tag{30}$$

This is identical to the Mark II form (Eq. (23)), discussed previously and illustrated in Figs. 3a-g. The Mark II and III forms therefore differ only in their unscaled constrained entropies $H_{q,i}^{c(II)}$ and $H_{q,i}^{c(III)}$, produced by differences in the structure of the system, as expressed by the $q$-partition function $Z_q^{(II)}$ or $Z_q^{(III)}$.

4. **Example System**

The utility of the "dismemberment" method, which allows the independent examination of the behaviour of each state, is best illustrated by an example. Consider a closed system subject to a single, energetic moment constraint (Eq. (9), (10) or (24) with $R=1$), with equispaced energy levels $F_{1i} = \varepsilon_i$, where $\langle F_1 \rangle$ or $\langle\langle F_1 \rangle\rangle = E$ = expected (mean) energy, and $\kappa_1 = \mu_1 = \nu_1 = \beta$ = Lagrangian multiplier. This may be interpreted physically as a quantum harmonic oscillator, or other everyday system such as steps in a staircase, floors of a car parking station, etc. Setting $\beta\varepsilon_i = i\chi$ and $\beta E = \hat{E}$, where $\chi$ is the normalised energy of the lowest energy level and $\hat{E}$ is a normalised expected energy, the three constrained Tsallis probability distributions (Eq. (14), (19), (27)) are:

$$p_i^{(I)} = \frac{1}{Z_q^{(I)}}\left(1 - i\chi(q-1)\right)^{1/(q-1)}, \quad i = 1,...,s, \tag{31}$$

$$p_i^{(II)} = \frac{1}{Z_q^{(II)}}\left(1 - i\chi(1-q)\right)^{1/(1-q)}, \quad i = 1,...,s, \tag{32}$$

$$p_i^{(III)} = \frac{1}{Z_q^{(III)}} \left(1 - (i\chi - \hat{E})(1-q)\right)^{1/(1-q)}, \quad i = 1,...,s. \tag{33}$$

The respective *q*-partition function is obtained by substitution of each probability function into the norm constraint (Eq. (2)); the expected energy by substitution into the relevant moment constraint (Eq. (9), (10) or (24)); and the partial constrained entropy function by substitution into Eq. (16), (23) or (30).

The above functions and other parameter values were calculated numerically for specified parameter settings for the Mark I, II and III cases. Note that for all constraint sets it is not possible to simultaneously fix $\chi$, $q$ and $\hat{E}$, since these parameters are linked through the moment constraint (Eq. (9), (10) or (24)). Solutions were therefore obtained by specifying $q$ and $\hat{E}$ (or $q$ and $\chi$), and recalculating $\chi$ (or $\hat{E}$) by iteration until convergence in the values of $Z_q$ and $\hat{E}$ (or $Z_q$ and $\chi$) was achieved. The $q=1$ values of $Z_q$ and $\chi$ (or $Z_q$ and $\hat{E}$) were used as initial seed values. In most analyses, the convergence criterion was a relative difference of $\pm 10^{-10}$, but in some cases this was relaxed to $\pm 10^{-9}$ to achieve convergence. Initially, the parameter settings used by Niven [9] for $q=1$ were adopted here ($s = 10$; seed $\chi = 1/3$, whence $\hat{E} = 1.052596469$ and $Z = 2.4375523757$ at $q=1$).

For the example given, the real-valued results for the Mark III case are illustrated in Figs. 4a-b, respectively for constant $\hat{E}$ and constant $\chi$. In both cases, the applicable range of $q$ was approximately $0.5 \leq q \leq 2.3$, with an additional outlier at $q=0$, outside of which $Z_q^{(III)}$ and either $\hat{E}$ or $\chi$ become complex. (Note the different definitions of *Z* (see Eq. (5)) and

$Z_q^{(III)}$ (see Eq. (27)), such that $Z_q^{(III)}=Z\exp\hat{E}$). In this particular example:

- When $\hat{E}$ is held constant, $Z_q^{(III)}$ decreases and χ increases with increasing $q$ over its valid range. The system therefore acts as if the spacing of energy levels increases with increasing $q$. Thus for $q<1$, energy associated with each state appears to be "lost", whilst for $q>1$, it appears to be "gained".

- When χ is fixed, $Z_q^{(III)}$ remains approximately constant (except for $q=0$) whilst $\hat{E}$ decreases with increasing $q$. The system therefore acts as if the normalised energy $\hat{E}$ decreases with increasing $q$. Thus for $q<1$, total energy appears to be "gained", whilst for $q>1$, it appears to be "lost".

Examining the probability distributions, plots of the function $S_{q,i}^{c(III)}$ for this example, for various values of $q$, are illustrated in Figs. 5a-e (for constant $\hat{E}$) and Figs. 6a-e (for constant χ). The $q=1$ plot (Figs. 5c and 6c) has been presented previously [9]. The stationary probability distributions in each case are included in Fig. 5f and 6f respectively. As evident:

- When either $\hat{E}$ or χ is fixed and $q=0$, the normalised constrained partial entropy (Figs. 5a, 6a) is horizontal for all states. The system is thus trapped in a single configuration, from which it cannot escape. For constant χ at $q=0$ (Figs. 6a, f), two of the probabilities are negative, i.e. this result is spurious.

- When either $\hat{E}$ or χ is fixed and $q>0$, as noted by Tsallis *et al.* [3], $q<1$ and $q>1$ respectively privilege the rare and frequent events. The effect of this behaviour on each individual state is clear in Figs. 5b-e and 6b-e. Compared to the *q*=1 case, at *q*=0.5 the individual $S_{q,i}^{c(III)}$ curves are more skewed to the right (i.e., privileging lower $p_i$), and also more evenly spaced. For *q*=1.5 and especially *q*=2.3, the individual $S_{q,i}^{c(III)}$ curves are skewed slightly to the left (privileging higher $p_i$). They are also much more unevenly spaced, such that the *i*=1 state is much more dominant.

- There are subtle differences between the results for fixed $\hat{E}$ or $\chi$. This includes the magnitude of the maximum entropy positions, which are higher (lower) for *q*<1 (*q*>1) when $\chi$ is fixed.

The plots in Figs. 5a-e and 6a-e clearly illustrate the utility of the "dismemberment" technique, as applied to Tsallisian statistical mechanics, in that it enables the independent examination of the behaviour of each state.

For the Mark III case, analysis of different parameter settings found that physical (non-complex) solutions were only obtained for $\chi \geq 1/3$ (either as a fixed value of $\chi$, or a seed value for constant $\hat{E}$). Since $\beta\varepsilon_i = i\chi$ and $\beta \propto 1/T$, where *T* is a dimensionless temperature, we see that $\chi \propto 1/T$; consequently, higher values of $\chi$ reflect "colder" conditions. Accordingly, the probability distributions become increasingly polarised with increasing $\chi$, with the first (*i*=1) state increasingly dominant. The minimum value of $\chi$ reveals the existence of a maximum temperature, a previously documented feature of Tsallis statistics [4].

With increasing χ, many of the trends with *q* shown in Figs. 4a-b change, such as between $Z_q$, $\hat{E}$ and χ. A summary of these trends is given in Table 1. In all cases, the trend in χ (for fixed $\hat{E}$) is opposite to that of $\hat{E}$ (for fixed χ). Relative to the *q*=1 position, virtual "losses" or "gains" in the energy of each state (when $\hat{E}$ is fixed) or total energy (when χ is fixed) are evident when *q*≠1. Such effects appear to be fundamental to Tsallisian statistical mechanics.

In the Mark III example, since *q*>0 (except at *q*=0) the constrained partial entropy is always concave. In contrast, both other Mark forms were less "well-behaved" for the example given. As an example, several plots of the function $H_{q,i}^{c(I)}$ for the Mark I case, for fixed $\hat{E}$, are illustrated in Fig. 7a-d, with the corresponding stationary probability distributions in Fig. 7e. As evident, the $H_{q,i}^{c(I)}$ functions "flip" from convex to concave curves as *q* crosses from *q*<0 to *q*>0. Note also the peculiar result for *q*=1.5 (Fig. 7d), in which the $H_{q,i}^{c(I)}$ functions diminish in magnitude from *i*=1 to 7, and then increase again from *i* = 8 to 10. This has a pronounced effect on the stationary probability distribution (Fig. 7e), which is convex. At least one peculiar probability distribution of this type was evident within each set of Mark I, II and III results (except for the Mark III case at χ=1/3), for both fixed $\hat{E}$ or χ. It is not clear whether such distributions are spurious, or reflect some peculiar, previously unreported probability distribution.

## 5. Conclusions

In this study, the constraining technique of Niven [9] is applied to the Tsallis entropy function $H_q$ [1], to determine its equivalent constrained forms when subject to the natural constraint and either the Mark I, II or III set of moment constraints. In each case, a constrained form of the Tsallis entropy function exists, which reduces to its Shannon equivalent (Eq (6)) when $q=1$. Since they are fully constrained (i.e., they are Lagrangian equations), each function can be "dismembered" to give its *partial* or *local* form, providing the means to examine each outcome or state independently of the other states. The Mark II and III constrained Tsallis entropy functions ($H_q^{c(II)}$ or $H_q^{c(III)}$) are found to depend not only on the probability ($p_i$), the stationary probability ($p_i^{(II)}$ or $p_i^{(III)}$) and $q$, but also on the respective $q$-partition function ($Z_q^{(II)}$ or $Z_q^{(III)}$). In contrast the Mark I form ($H_q^{c(I)}$) depends only on $p_i$, $p_i^{(I)}$ and $q$. The Mark II and III forms therefore depend on the overall structure of the system, as expressed by the $q$-partition function, and not just on the individual properties of state $i$. Such behaviour accords with what we know of non-extensive statistical mechanics.

The partial forms of all three functions are examined in general, and by consideration of an example of a system with equispaced energy levels, providing a visual illustration of several features of Tsallisian statistical mechanics.

**Table 1: Trends of $Z_q^{(III)}$, χ and $\hat{E}$ with increasing *q*, for different values of χ, for the example considered in §4 (Mark III case).**

| Seed χ | Fixed $\hat{E}$ | | Fixed χ | |
|---|---|---|---|---|
| | $Z_q^{(III)}$ | χ | $Z_q^{(III)}$ | $\hat{E}$ |
| 1/3 | ↓ | ↑ | ~ Constant | ↓ |
| 3/10 | ↓ | ↑ | Undulating | ↓ |
| ½ | Convex | ↑ | ↑ | ↓ |
| 1 | ↑ | ↑ | ↑ | ↓ |
| 2 | ↑ | Convex | ↑ | Concave |
| 3 | ↑ | ↓ | ↑ | ↑ |
| 5 | ↑ | ↓ | ↑ | ↑ |
| 10 | ↑ | ↓ | ↑ | ↑ |
| 25 | ↑ | Slight ↓ | ↑ | Slight ↑ |
| 50 | Slight ↑ | ~ Constant | Slight ↑ | ~ Constant |

**FIGURE LIST**

Fig. 1: Three-dimensional plot of the extensive constrained partial entropy function $H_i^c$ (Eq. (7)) against $p_i$ and $p_i^*$ (after [9]).

Fig. 2: Three-dimensional plots of $H_{q,i}^{c(I)}$ (Eq. (16)) against $p_i$ and $p_i^{(I)}$, for various values of $q$. (The $q=1$ case is shown in Fig. 1.)

Fig. 3: Three-dimensional plots of $S_{q,i}^{c(II)}$ (Eq. (23)) against $p_i$ and $p_i^{(II)}$ (identical to $S_{q,i}^{c(III)}$ (Eq. (30)) against $p_i$ and $p_i^{(III)}$), for various values of $q$. (The $q=1$ case is shown in Fig. 1.)

Fig. 4: Numerical results for the example of §4: (a) plots of $Z_q^{(III)}$ and $\chi$ against $q$ (assuming $\hat{E} = 1.052596469$, $s = 10$); and (b) plots of $Z_q^{(III)}$ and $\hat{E}$ against $q$ (assuming $\chi = 1/3$, $s = 10$). Note that for $q=1$, for which $Z$ is normally used (see Tables 1-2), $Z_q^{(III)} = Z \exp \hat{E}$ is represented by an open diamond.

Fig. 5: (a-e) Plots of $S_{q,i}^{c(III)}$ (Eq. (23)) against $p_i$ for the example of §4, assuming constant $\hat{E} = 1.052596469$, for several values of $q$; and (f) the corresponding stationary probability distributions $p_i^{(III)}$ against $i$. Plot (c) is after [9].

Fig. 6: (a-e) Plots of $S_{q,i}^{c(III)}$ (Eq. (23)) against $p_i$ for the example of §4, assuming constant $\chi = 1/3$, for several values of $q$; and (f) the corresponding stationary probability distributions $p_i^{(III)}$ against $i$. Plot (c) is after [9].

Fig. 7: (a-d) Plots of $H_{q,i}^{c(I)}$ (Eq.(16)) against $p_i$ for the example of §4, assuming

constant $\hat{E} = 1.052596469$, for several values of $q$; and (e) the corresponding stationary probability distributions $p_i^{(I)}$ against $i$. For $q=0$ (not shown), $H_{q,i}^{c(I)}=1, \forall i$.

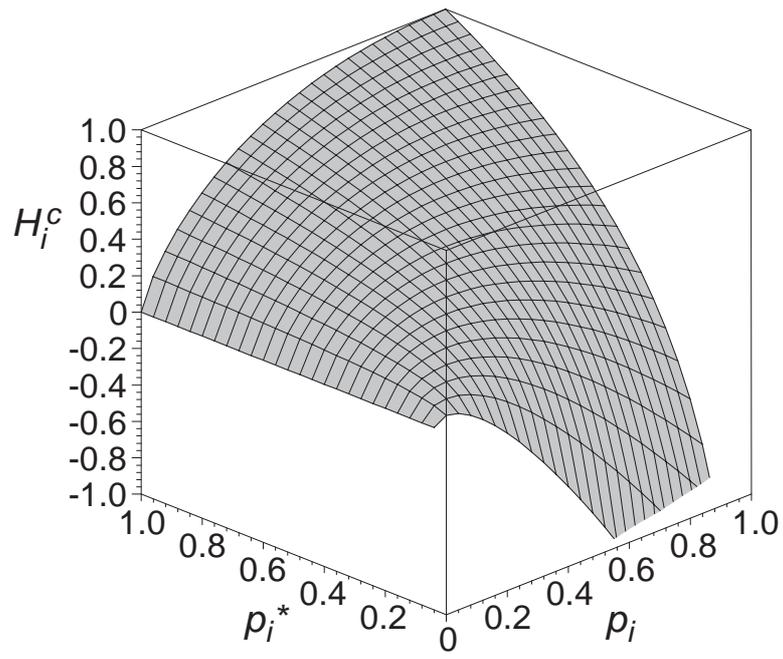

FIGURE 1

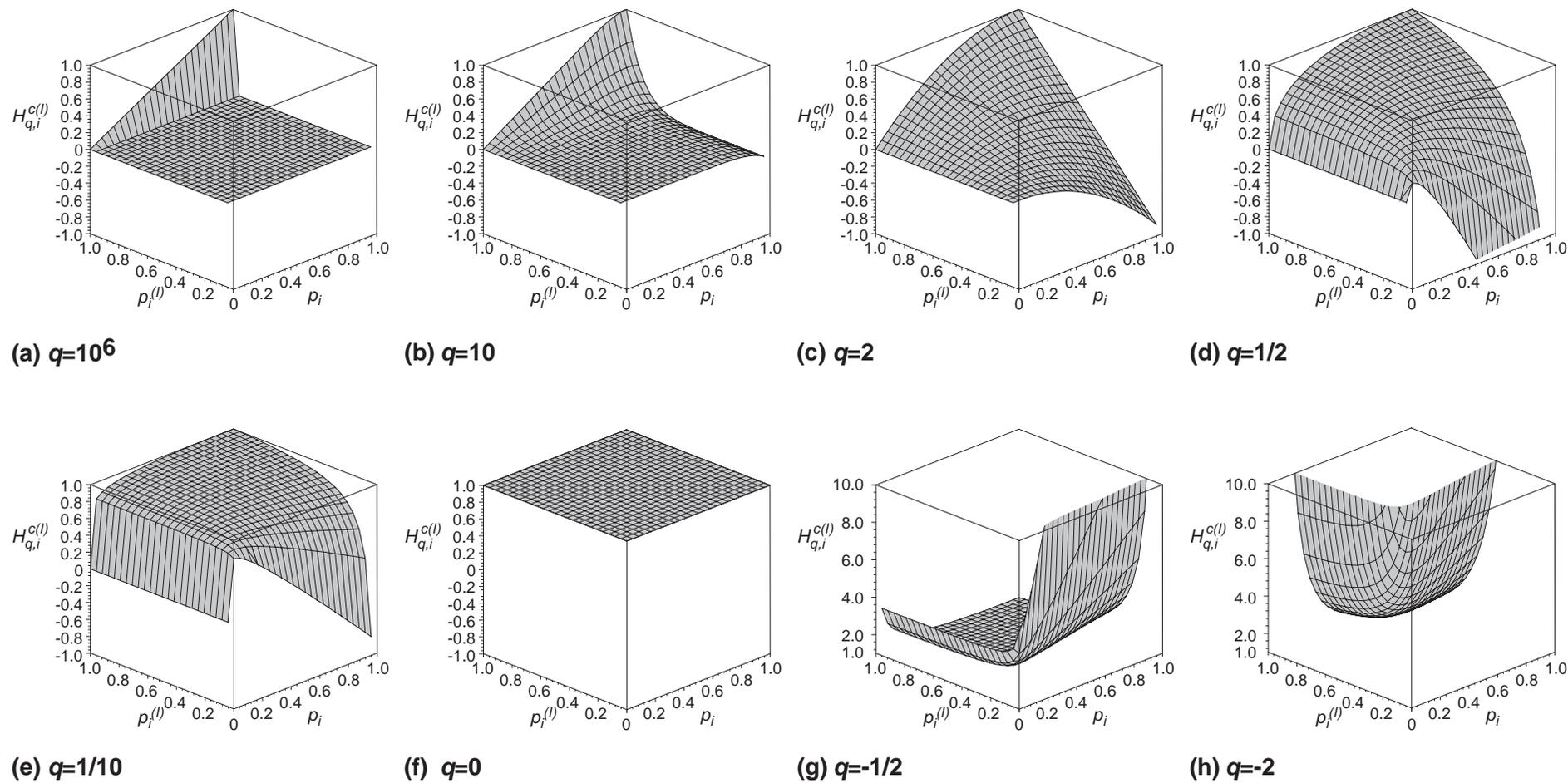

(a) $q=10^6$    (b) $q=10$    (c) $q=2$    (d) $q=1/2$

(e) $q=1/10$    (f) $q=0$    (g) $q=-1/2$    (h) $q=-2$

FIGURE 2

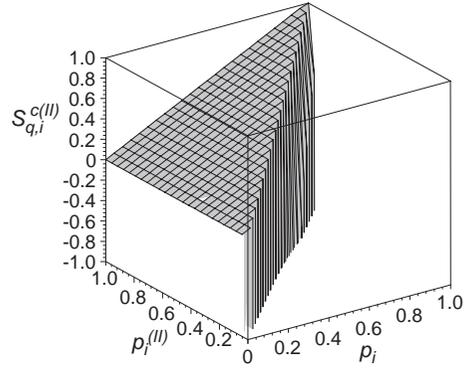

**(a)** $q=100$

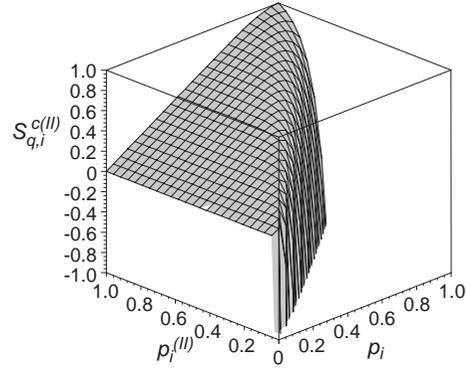

**(b)** $q=10$

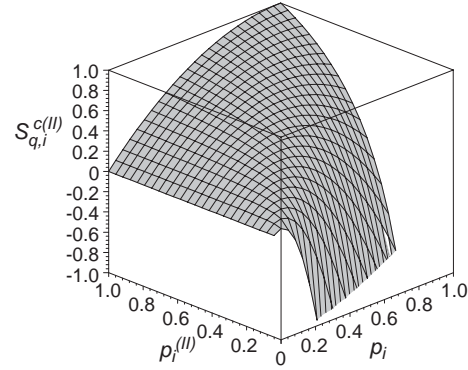

**(c)** $q=2$

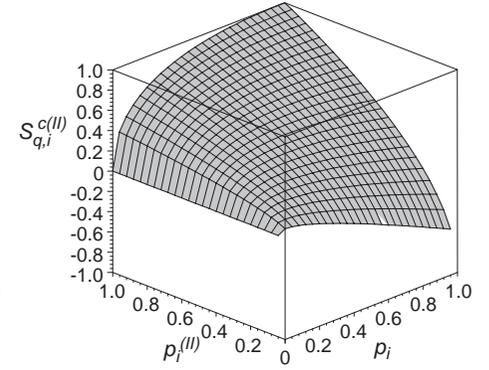

**(d)** $q=1/2$

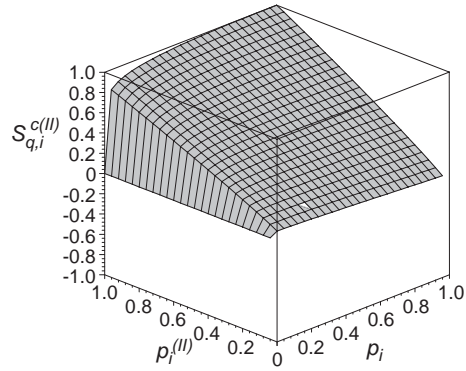

**(e)** $q=1/10$

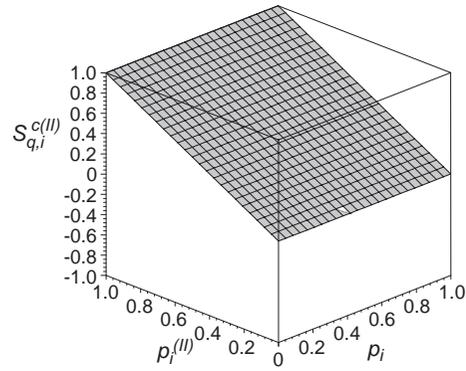

**(f)** $q=0$

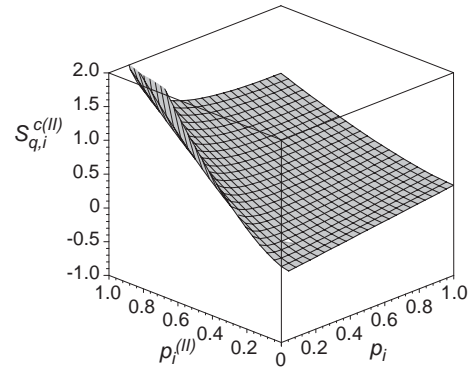

**(g)** $q=-1/2$

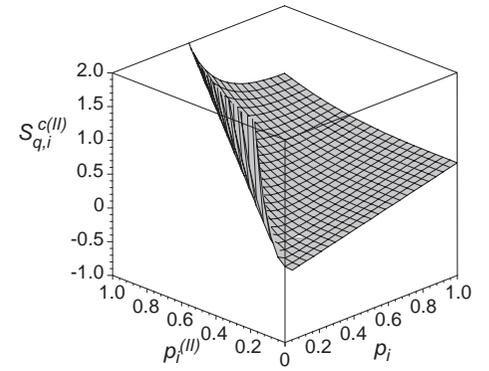

**(h)** $q=-2$

FIGURE 3

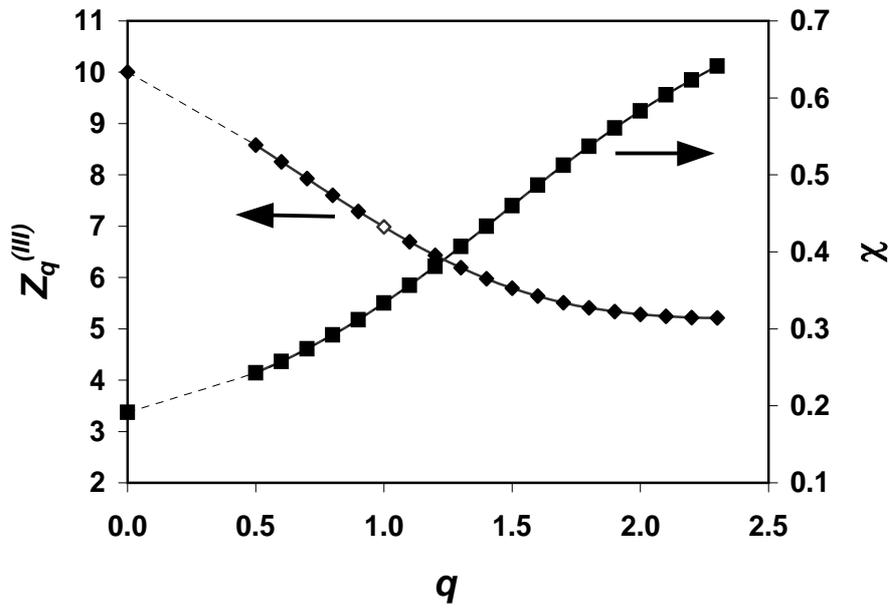

**(a)**

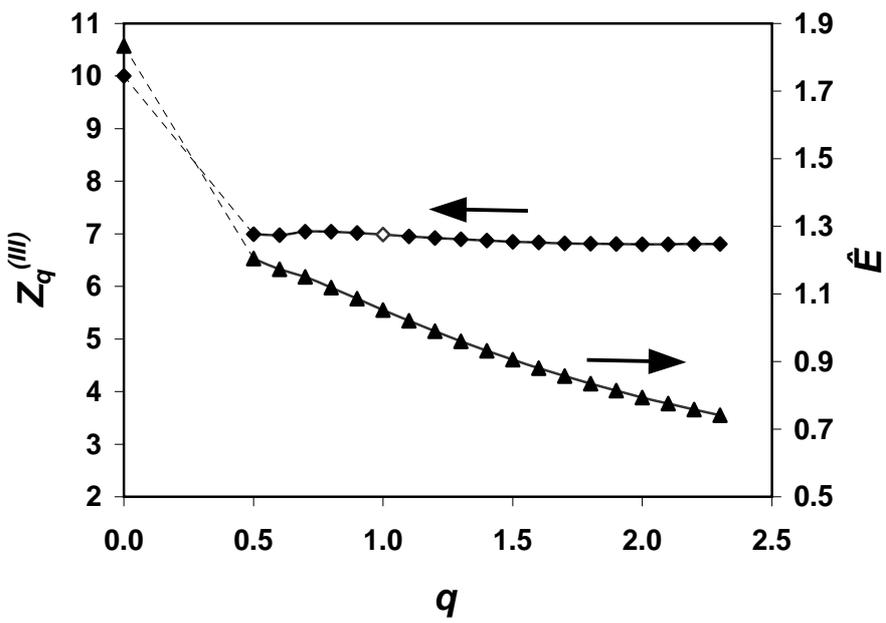

**(b)**

FIGURE 4

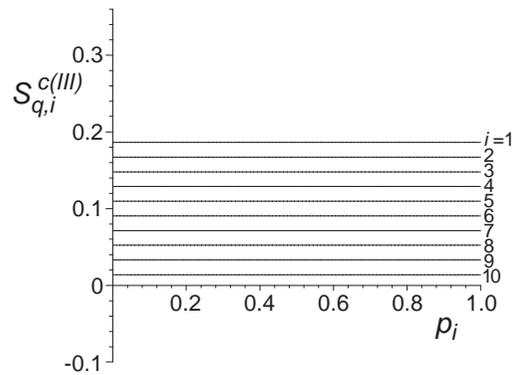 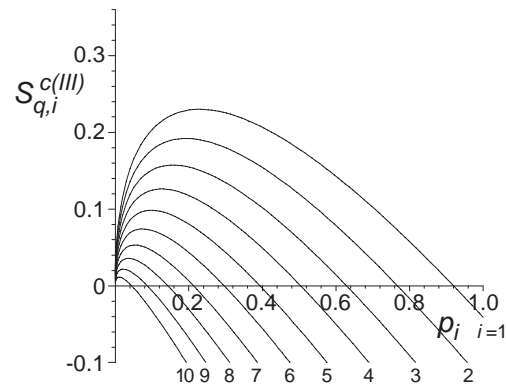 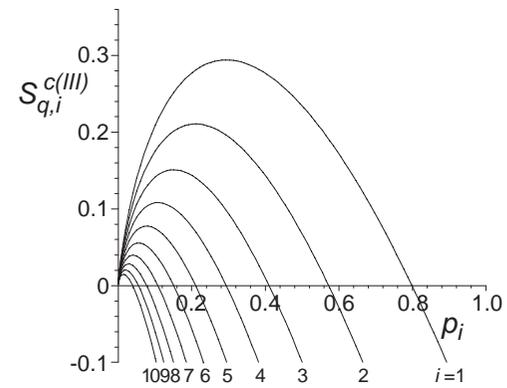

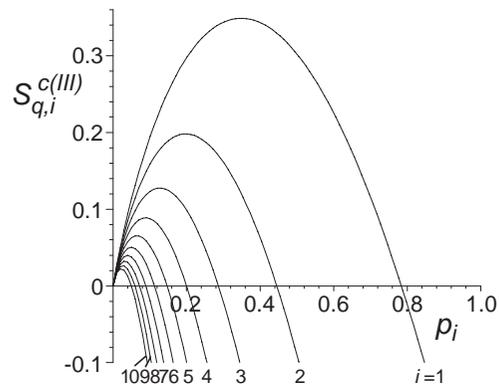 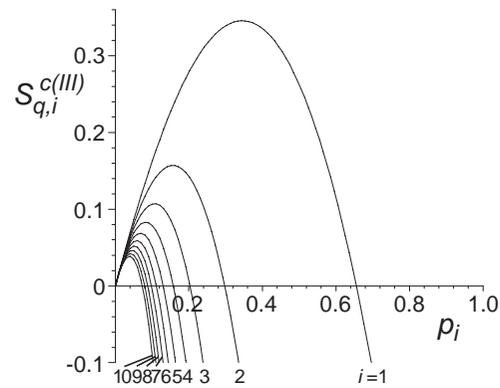 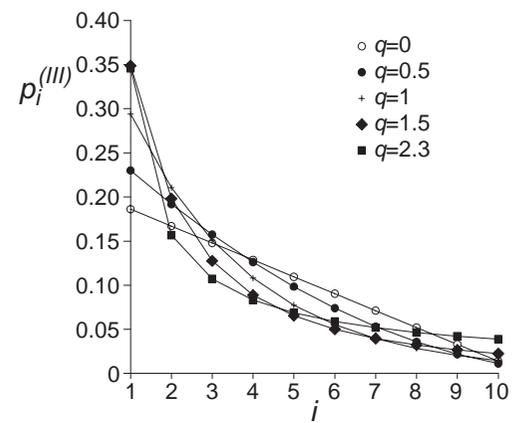

FIGURE 5

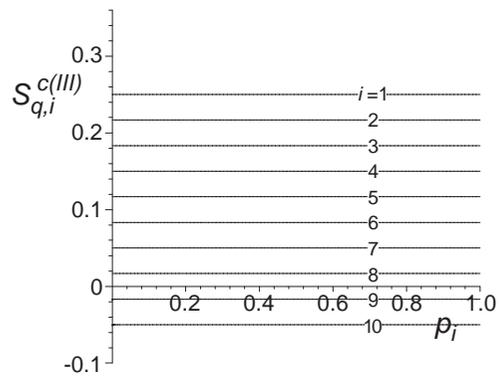
**(a)** $q=0$

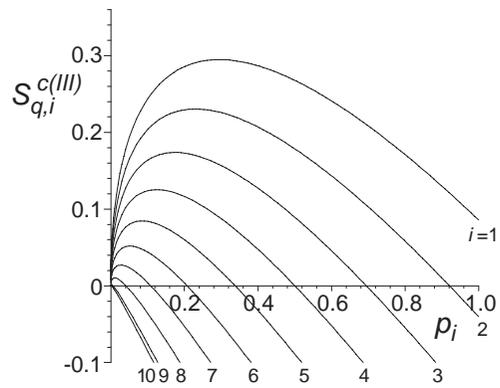
**(b)** $q=0.5$

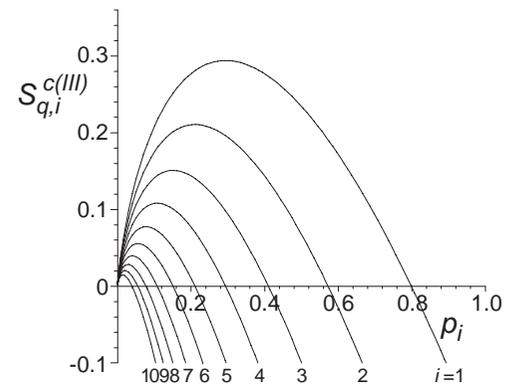
**(c)** $q=1$

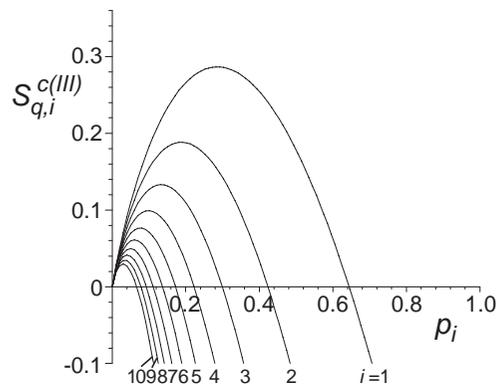
**(d)** $q=1.5$

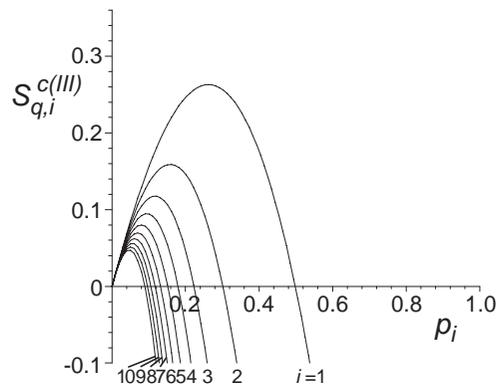
**(e)** $q=2.3$

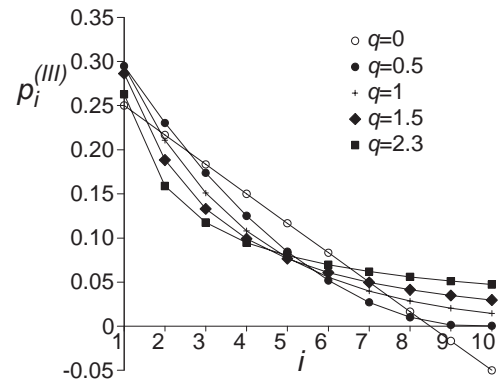
**(f)**

FIGURE 6

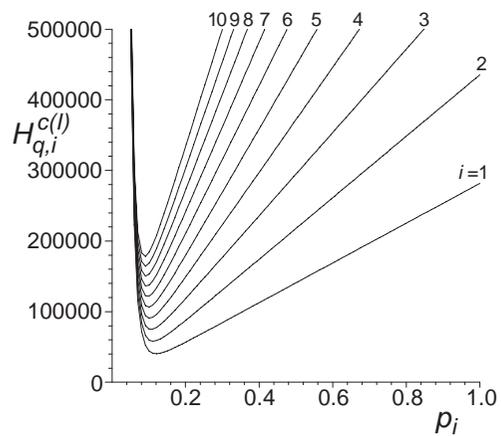

**(a)** $q$=-5.0

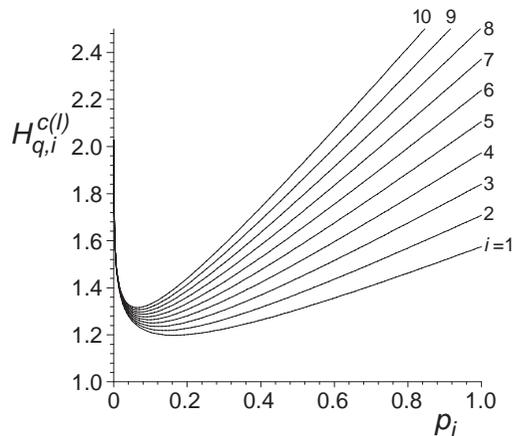

**(b)** $q$=-0.1

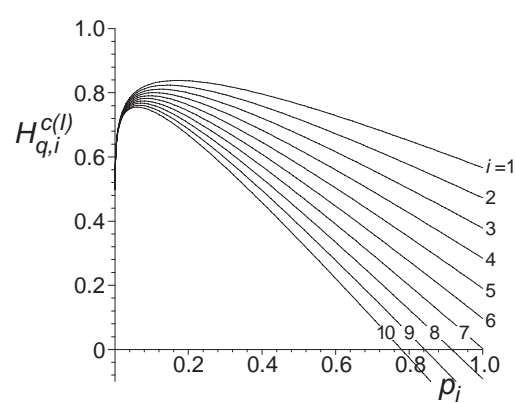

**(c)** $q$=0.1

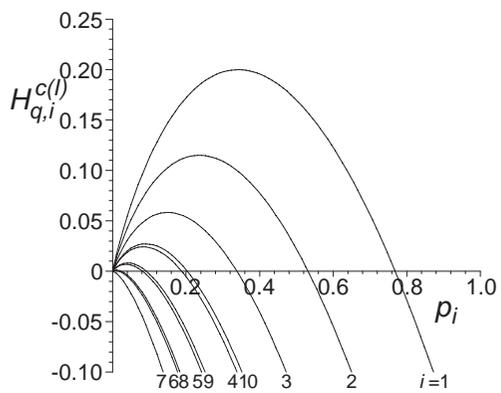

**(d)** $q$=1.5

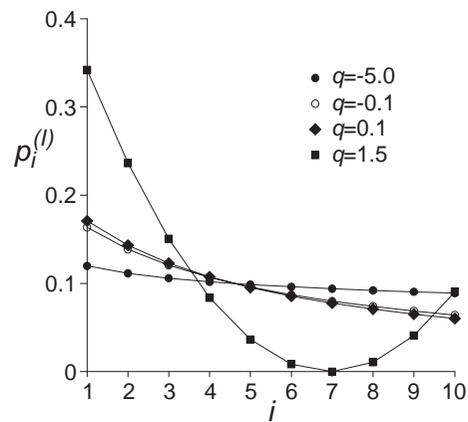

**(e)**

FIGURE 7

**30**